\documentclass[aps,preprint,preprintnumbers,12pt,floatfix]{revtex4}
\usepackage{epsfig,psfrag}
\usepackage{latexsym}
\def\nn{\nonumber}
\def\tg{\tilde{\gamma}}

\newcommand{\be}{\begin{equation}}
\newcommand{\ee}{\end{equation}\noindent}
\newcommand{\bear}{\begin{eqnarray}}
\newcommand{\ear}{\end{eqnarray}\noindent}

\newcommand{\slD}{\raise.15ex\hbox{$/$}\kern-.57em\hbox{$D$}}
\newcommand{\slpartial}{\raise.15ex\hbox{$/$}\kern-.57em\hbox{$\partial$}}
\newcommand{\slG}{{{\dot G}\!\!\!\! \raise.15ex\hbox {/}}}

\def\mn{\mu\nu}

\def\non{\nonumber}
\def\beqn*{\begin{eqnarray*}}
\def\eqn*{\end{eqnarray*}}

\def\square{\kern1pt\vbox{\hrule height 1.2pt\hbox{\vrule width 1.2pt
   \hskip 3pt\vbox{\vskip 6pt}\hskip 3pt\vrule width 0.6pt}
   \hrule height 0.6pt}\kern1pt}
\def\slash#1{#1\!\!\!\raise.15ex\hbox {/}}

\def\dps{\displaystyle}

\def\half{{1\over 2}}

\def\e{\mbox{e}}

\def\4piTD{{(4\pi T)}^{-{D\over 2}}}
\def\4piT4{{(4\pi T)}^{-2}}

\def\Tintm4{{\dps\int_{0}^{\infty}}{dT\over T}\,e^{-m^2T}
    {(4\pi T)}^{-2}}
\def\Tintm{{\dps\int_{0}^{\infty}}{dT\over T}\,e^{-m^2T}}

\def\tr{{\rm tr}\,}

\def\be{\begin{equation}}\def\ee{\end{equation}}
\def\bea{\begin{eqnarray}}\def\eea{\end{eqnarray}}
\def\ba{\begin{array}}\def\ea{\end{array}}
\def\bea{\begin{eqnarray}}\def\barr{\begin{array}}\def\earr{\end{array}}
\def\eea{\end{eqnarray}}
\begin{document} 
\newcommand{\ho}[1]{$\, ^{#1}$}
\newcommand{\hoch}[1]{$\, ^{#1}$}

\title{Worldline Instantons and Pair Production in Inhomogeneous Fields}

\author{Gerald V. Dunne}
\affiliation{Department of Physics,
University of Connecticut, Storrs, CT 06269-3046, USA}
\author{Christian Schubert\footnote{Address after 9/2005: Instituto de F\'{\i}sica y Matem\'aticas
\\
Universidad Michoacana de San Nicol\'as de Hidalgo\\
Ciudad Universitaria, Edificio C-3, Apdo. Postal 2-82\\
58040, Morelia, Michoac\'an, Mexico}
}
\affiliation{Department of Physics and Geology,
University of Texas Pan American, Edinburg, TX 78541-2999, USA}

\begin{abstract}
We show how to do semiclassical nonperturbative computations within the worldline approach to quantum field theory using ``worldline instantons''. These worldline instantons are classical solutions to the Euclidean worldline loop equations of motion, and are closed spacetime loops parametrized by the proper-time. Specifically, we compute the imaginary part of the one loop effective action in scalar QED using ``worldline instantons'', for a wide class of inhomogeneous electric field backgrounds. We treat both time dependent and space dependent electric fields, and note that temporal inhomogeneities tend to shrink the instanton loops, while spatial inhomogeneities tend to expand them. This corresponds to temporal inhomogeneities tending to enhance local pair production, with spatial inhomogeneities tending to suppress local pair production. We also show how the worldline instanton technique extends to spinor QED.

\end{abstract}


\maketitle

\vspace{20pt}

\section{Introduction}

The worldline path integral 
formulation of quantum field theory provides a powerful computational approach to both perturbative and nonperturbative phenomena. Originally due to Feynman \cite{feynman}, it has in recent
years attracted renewed attention due to its relation to string theory 
methods \cite{halpern,bern,polyakov,strassler} (see \cite{csreview} for a review). Among other advantages, 
it provides a particularly efficient way of taking constant external fields into account nonperturbatively
\cite{ss1,adlsch,shaisultanov,rescsc,bassch}. Very recently, numerical techniques have been
developed for the calculation of worldline path integrals \cite{gies,schmidt} which raise the prospect of doing calculations which are nonperturbative in the coupling \cite{gisava}. 
 
One of the most interesting nonperturbative phemonena is vacuum pair production \cite{sauter,heisenberg,schwinger}, which has applications in many fields of physics \cite{ringwald}, ranging from particle and nuclear physics \cite{nussinov,dima} to cosmology \cite{starobinsky,parker,hawking}. 
Already in 1981 Affleck et al \cite{affleck} studied pair production in a constant field in scalar QED
by applying instanton techniques to Feynman's worldline path integral. As is well-known, the QED
pair creation by an external field can be concisely described in terms of the imaginary part of
the effective lagrangian. 
In scalar QED at one loop and for a constant field this imaginary part is given by Schwinger's 
formula \cite{schwinger},
\bear
{\rm Im}\,{\cal L}[E] 
&=&
\frac{e^2 E^2}{16\pi^3}\, \sum_{n=1}^\infty \frac{(-1)^{n-1}}{n^2}
\,\exp\left[-\frac{m^2 \pi n}{e E}\right]
\label{L1scalim}
\end{eqnarray}
This expression is clearly non-perturbative in terms of
the field. The $n^{\rm th}$ term in the sum directly relates to the probability for the coherent
production of $n$ pairs by the field \cite{schwinger,lebedev}. Usually one derives the formula (\ref{L1scalim}) by a proper treatment of the poles appearing in the standard integral representation of the 
one-loop Euler-Heisenberg lagrangian 
(see, e.g., \cite{ditreubook,jentschura}). Affleck et al. showed in \cite{affleck} that the same formula can be obtained in the spirit of instanton physics by a stationary phase appromixation of the corresponding worldline path integral. 
In this paper we further develop this semiclassical approach to worldline computations of pair production in QED, based on special stationary worldline loops which we call worldline instantons. 

Worldline instantons also provide the worldline formulation of the conventional field theoretic WKB computations of Brezin et al \cite{brezin} and Popov et al \cite{popov}, which were in turn based on the pioneering ionization studies of Keldysh \cite{keldysh}. We note that Kim and Page \cite{kimpage} have discussed pair production in the WKB approach using the language of quantum mechanical instantons. These quantum mechanical instantons are not the same as our "worldline instantons", which are instantons in the proper-time, rather than in the imaginary time of quantum mechanical tunneling computations. Although both approaches
are related, from the point of view of relativistic field theory the use of proper-time seems more
natural. The {\it numerical} worldline approach has recently also been applied directly to pair production phenomena \cite{giesklingmuller}.

In Section \ref{phase} we discuss the general idea of  worldline instantons as a semiclassical approximation to the worldline effective action. In Sections \ref{time} and \ref{space} we present some explicit examples for electric fields that are time and space dependent, respectively. Section \ref{spinor} explains how worldline instantons appear in the imaginary part of the spinor effective action, and Section \ref{conclusions} contains a unified summary and outlines future work.

\section{Stationary Phase Approximation in the Worldline Effective Action}
\label{phase}

The Euclidean one-loop effective action for a scalar particle in an abelian gauge background $A_\mu$ is given
by the worldline path integral expression \cite{feynman,csreview}
\bear
\Gamma [A] =
\int_0^{\infty}\frac{dT}{T}\, \e^{-m^2T}
\int_{x(T)=x(0)} {\mathcal D}x 
\, {\rm exp}\left[-\int_0^Td\tau 
\left(\frac{\dot x^2}{4} +i e A\cdot \dot x \right)\right]
\label{PI}
\ear
Here the functional integral $\int {\mathcal D}x$ is over all closed spacetime paths $x^\mu(\tau)$ which are periodic in the proper-time parameter $\tau$, with period $T$. The effective action $\Gamma[A]$ is a functional of the classical background field $A_\mu(x)$, which is a given function of the space-time 
coordinates.
Rescaling $\tau = Tu$, and $T$ by $m^2$, the effective action may be expressed as
\bear
\Gamma [A] =
\int_0^{\infty}\frac{dT}{T}\, \e^{-T}
\int_{x(1)=x(0)} {\mathcal D}x 
\, {\rm exp}\left[-\left(\frac{m^2}{4T}\int_0^1du \,
\dot x^2 +i e\int_0^1du \, A\cdot \dot x 
\right)\right]
\label{PIscale}
\ear
where the functional integral $\int {\mathcal D}x$ is now over all closed spacetime paths $x^\mu(u)$ with period $1$.
With this rescaling we can perform the proper-time integral explicitly, leading to 
\bear
\Gamma [A] &=&
2 \int_{x(1)=x(0)} {\mathcal D}x 
\, K_0\left(m\sqrt{\int_0^1 du\,\dot{x}^2}\right)\, {\rm exp}\left[- i e\int_0^1du\, A\cdot \dot x 
\right]\\
&\simeq& \sqrt{\frac{2\pi}{m}}\, \int {\mathcal D}x \, \frac{1}{\left(\int_0^1 du\, \dot x^2\right)^{1/4}}\, 
{\rm exp}\left[-\left(m\sqrt{\int_0^1 du\, \dot x^2} 
+i e\int_0^1 du\, A\cdot \dot x
\right)\right]
\label{bessel}
\ear
Here we have used the approximation 
\bear
m\sqrt{\int_0^1 du\, \dot{x}^2}\gg 1 \quad ,
\label{large}
\ear
which, as we will see later, corresponds to a weak field condition. Alternatively, the second line in (\ref{bessel}) could be obtained by evaluating the $T$ integral in (\ref{PIscale}) using Laplace's method, with critical point \bear
T_0=\frac{m}{2}\sqrt{\int_0^1 du\, \dot{x}^2} \,\,\gg 1
\label{critical}
\ear

The functional integral remaining in the effective action expression (\ref{bessel}) may be approximated by a functional stationary phase approximation. The new, nonlocal, worldline ``action''  \cite{affleck},
\bear
S = m\sqrt{\int_0^1 du\, \dot x^2} + i e \int_0^1duA\cdot \dot x
\label{defS}
\ear
is stationary if the path $x_\alpha(u)$ satisfies 
\bear
m{\ddot x_{\mu}\over \sqrt{\int_0^1 du\, \dot x^2}} &=& i e F_{\mn}\dot x_{\nu}
\label{statcond}
\ear
We call a periodic solution $x_\mu(u)$ to (\ref{statcond}) a ``worldline instanton''. 
Recall that $F_{\mu\nu}=\partial_\mu A_\nu-\partial_\nu A_\mu$ is in general a function of $x_\alpha(u)$, so the stationarity condition (\ref{statcond}) is a highly nontrivial set of coupled nonlinear (ordinary) differential equations. Note, however, that by contracting (\ref{statcond}) with $\dot{x}_\mu$ we learn that, for any $F_{\mu\nu}(x)$, the stationary instanton paths satisfy 
\bear
\dot{x}^2={\rm constant}\equiv a^2
\label{c2}
\ear
Thus, the condition (\ref{large}) can be written as $m a\gg 1$.

In a background electric field the fluctuations about the worldline instanton paths lead to an imaginary part in the effective action $\Gamma[A]$, and the leading behavior is
\bear
{\rm Im}\,\Gamma[A]\sim e^{-S_0}\quad ,
\label{leading}
\ear
where $S_0$ is the worldline action (\ref{defS}) evaluated on the worldline instanton. This imaginary part of the effective action gives the pair production rate \cite{heisenberg,schwinger}. 
In a future paper \cite{ds-flucs} we discuss the subleading prefactor contributions to (\ref{leading}).

\section{Temporally Inhomogeneous Electric Fields}
\label{time}

In this Section we consider a class of time dependent background fields for which we are able to find the stationary instanton paths explicitly. Consider a classical electromagnetic background which in Minkowski space corresponds to a time-dependent electric field pointing in the $x_3$ direction. In Euclidean space, we choose a gauge in which the only nonzero component of the gauge field is $A_3$, and it is a function only of $x_4$:
\bear
A_3=A_3(x_4) \quad ; \quad A_\mu=0\,\,\,{\rm for}\,\, \mu\neq 3
\label{gauge}
\ear
Since $F_{\mu 1}=F_{\mu 2}=0$, the stationarity conditions (\ref{statcond}) imply that 
\bear
\ddot{x}_1=\ddot{x}_2=0 \quad \Rightarrow \quad \dot{x}_1= {\rm constant}\quad, \quad \dot{x}_2={\rm constant}
\label{x12}
\ear
For $x_1(u)$ and $x_2(u)$ to be periodic, we require $\dot{x}_1=\dot{x}_2=0$. Hence, by (\ref{c2}), we have
\bear
a^2=\dot{x}_3^2+\dot{x}_4^2
\label{c}
\ear
Thus, the stationarity conditions (\ref{statcond}) reduce to two coupled (in general nonlinear) equations for $x_3(u)$ and $x_4(u)$:
\bear
\ddot{x}_{3} &=& \frac{i e a}{m}\, F_{34}\, \dot{x}_{4}
\label{coupled1}
\\
\ddot{x}_{4}&=& -\frac{i e a}{m}\, F_{34}\, \dot{x}_{3}
\label{coupled2}
\ear
For background fields of the form (\ref{gauge}), the stationary conditions (\ref{coupled1},\ref{coupled2}) can be integrated as follows. First, (\ref{coupled1}) implies, using the periodicity of the solutions, that
\bear
\dot{x}_3=-\frac{i e a}{m}\, A_3(x_4) 
\label{x3}
\ear
Given this solution for $\dot{x}_3$, we can use (\ref{c}) to write
\bear
| \dot{x}_4 |=a\, \sqrt{1+\left(\frac{e\, A_3(x_4)}{m}\right)^2}
\label{x4}
\ear
For certain functions $A_3(x_4)$, this equation (\ref{x4}) can be explicitly integrated to find the unique periodic solution for $x_4(u)$. (Alternatively, it can be integrated numerically if no explicit closed-form solution is available). Given this solution, we can then integrate (\ref{x3}) to obtain the corresponding unique periodic solution for $x_3(u)$.

Also, note that for background fields of the form (\ref{gauge}),
the worldline action (\ref{defS})  can be re-written in a simpler form when evaluated on solutions of the stationary conditions (\ref{coupled1},\ref{coupled2}):
\bear
S_0&=&m\sqrt{ \int_0^1du\, \dot{x}^2}+i e\int_0^1 du\,\frac{dx_3}{du}\, A_3(x_4)\non\\
&=&m\, a+i e \int_0^1 du\, x_3\, F_{34}\, \dot{x}_4\non\\
&=& \frac{m}{a}\int_0^1du\, (\dot{x}_4)^2
\label{s0}
\ear
where we have integrated by parts, making use of the periodicity of the solutions in $u$, the stationarity equation (\ref{coupled1}), and the solution property
(\ref{c}). In the following subsections we present some explicit examples.

\subsection{Constant electric background}
\label{tconstant}

For a constant electric background, of magnitude $E$, the Euclidean gauge field is $A_3(x_4)=-i\, E\, x_4$. Then the stationary solution for $x_4(u)$ is obtained by integrating (\ref{x4}):
\bear
\left |\frac{d x_4}{du}\right |=a\sqrt{1-\left(\frac{e E}{m}\right)^2 x_4^2}
\ear
which has the solution 
\bear
x_4(u)=\frac{m}{e E}\, \sin\left(\frac{e E a}{m}\, u\right)
\label{x4sol}
\ear
The stationary solution for $x_3$ is then obtained by integrating (\ref{x3}),
\bear 
\frac{dx_3}{du}=-a\, \sin\left(\frac{e E a}{m}\, u\right)
\ear
which has the solution
\bear
x_3(u)=\frac{m}{e E}\,\cos\left(\frac{e E a}{m}\, u\right)
\label{x3sol}
\ear
These solutions for $x_3(u)$ and $x_4(u)$ are periodic with period $1$ provided the constant $a$ satisfies
\bear
a=\frac{m}{e E}\,2 n\, \pi\quad , \quad n\in {\bf Z}^+
\label{aperiod}
\ear
Having determined the constant $a$, we can now understand the physical meaning of the approximation (\ref{large}), which given (\ref{c2}) is equivalent to $ma\gg 1$. The condition
(\ref{large}) becomes
\bear
\frac{m^2}{e E} 2 \pi n \gg 1
\label{weakfield1}
\eea
This must be true for all n, so this is a weak-field condition. With physical constants reinstated it reads: $E\ll \frac{2\pi m^2 c^3}{e \hbar}\sim 10^{16} V/{\rm cm}$, which is strongly satisfied for experimentally accessible electric fields \cite{ringwald}.

\begin{figure}[ht]
\centerline{\includegraphics[scale=0.9]{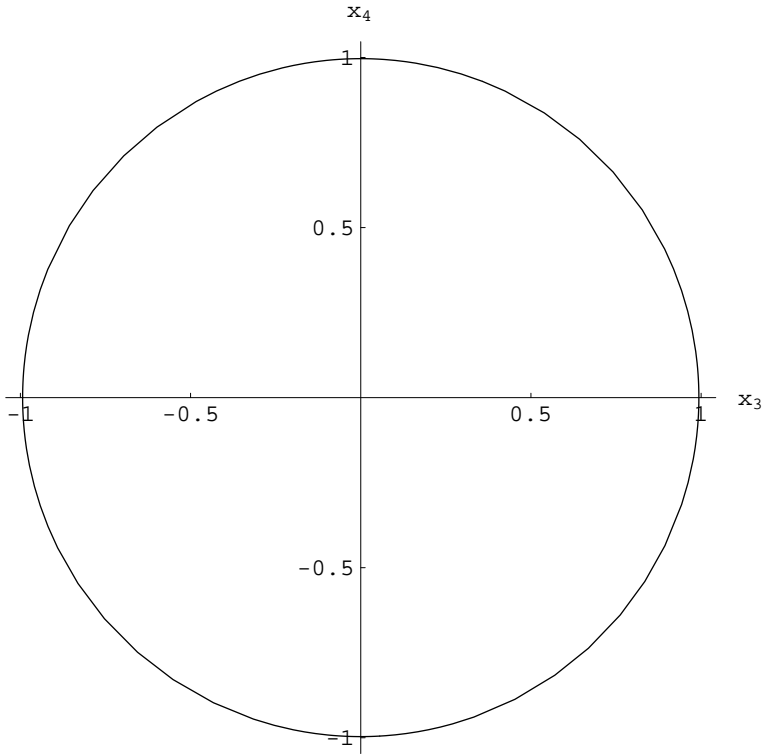}}
\begin{picture}(40,20)(60,-30)
\end{picture}
\caption{Parametric plot of the stationary worldline instanton paths in the $(x_3, x_4)$ plane for the case of a constant electric field of strength $E$. The paths are circular and the radius has been expressed in units of $\frac{m}{eE}$.}
\label{fig1}
\end{figure}

The stationary worldline instanton paths trace out circles (see Figure \ref{fig1}) of radius $\frac{m}{eE}$ \cite{affleck}:
\bear
x_3(u)=\frac{m}{eE}\,\cos(2 n \pi u)\quad , \quad x_4(u)=\frac{m}{eE}\,\sin(2 n \pi u)
\label{circle}
\ear
The integer $n$ simply counts the number of times the closed path is traversed.
The corresponding instanton action (\ref{s0}) is
\bear
S_0&=&\frac{m}{a}\, \left(\frac{2 n \pi m}{e E}\right)^2 \int_0^1 \cos(2 n \pi u)^2 \nn\\
&=&n\,\frac{m^2\pi}{eE}
\label{action-constant}
\ear
Thus we recover the $n$-instanton contribution to the imaginary part of the effective lagrangian, eq. (\ref{L1scalim}). 

\subsection{Single-pulse time dependent  electric background : $E(t)=E\, {\rm sech}^2(\omega\,t)$.}
\label{tsech}

The Minkowski time-dependent electric field $E(t)=E\, {\rm sech}^2(\omega\,t)$ corresponds to the Euclidean space gauge field
\bear 
A_3(x_4)=- i\,\frac{E}{\omega}\,\tan(\omega\, x_4) \quad
\ear
This case is soluble in the sense that the imaginary part of the effective action can be expressed as a momentum integral of a known function \cite{nn}. It has also been analyzed using WKB \cite{popov}. It is useful to introduce the ``adiabaticity parameter'', $\gamma$, (motivated by Keldysh's pioneering work on ionization in time dependent fields \cite{keldysh}), defined by 
\bear
\gamma\equiv \frac{m\omega}{eE}
\label{ad}
\ear
Then the stationary $x_4(u)$ is determined by integrating (\ref{x4}),
\bear
du=\frac{1}{a}\frac{d x_4}{\sqrt{1-\frac{1}{\gamma^2}\tan^2(\omega \, x_4)}}
\ear
which has the solution
\bear
x_4=\frac{1}{\omega}\, \arcsin\left[\frac{\gamma}{\sqrt{1+\gamma^2}}\, \sin\left(\frac{\sqrt{1+\gamma^2}}{\gamma}\, \omega \, a\, u\right)\right]
\label{x4sech}
\ear
Then, given this solution for $x_4(u)$, the stationary $x_3(u)$ is determined by integrating (\ref{x3})
\bear
\frac{d x_3}{du}=- a\, \frac{\sin\left(\frac{\sqrt{1+\gamma^2}}{\gamma}\, \omega \, a\, u\right)}{\sqrt{1+\gamma^2 \cos^2\left(\frac{\sqrt{1+\gamma^2}}{\gamma}\, \omega \, a\, u\right)}}
\ear
This has the solution
\bear
x_3=\frac{1}{\omega}\,\frac{1}{\sqrt{1+\gamma^2}} \, {\rm arcsinh}\left[\gamma\, \cos\left(\frac{\sqrt{1+\gamma^2}}{\gamma}\, \omega \, a\, u\right)\right]
\label{x3sech}
\ear
It is easy to verify that $\dot{x}_3^2+\dot{x}_4^2=a^2$ is satisfied.
Demanding that the solutions (\ref{x4sech}) and (\ref{x3sech}) be periodic in $u$, with period 1, fixes the constant $a$ to be
\bear
a=\frac{\gamma}{\omega\, \sqrt{1+\gamma^2}}\,2\pi  n\quad , \quad n\in {\bf Z}^+
\label{asech}
\ear
In this inhomogeneous case, the approximation condition (\ref{large}) becomes
\bear
\frac{m^2}{e E} \frac{1}{\sqrt{1+\gamma^2}} 2 \pi n \gg 1
\eea
As in the constant field case (\ref{weakfield1}), this is a weak-field condition, although it also includes the adiabaticity parameter $\gamma$, which cannot be too large for a given peak field $E$.

The periodic stationary worldline instanton paths are:
\bear
x_3(u)&=&\frac{m}{e E}\,\frac{1}{\gamma \sqrt{1+\gamma^2}} \, {\rm arcsinh}\left[\gamma\, \cos\left(2 n \pi u\right)\right]\nonumber \\
x_4(u)&=&\frac{m}{e E}\, \frac{1}{\gamma}\, \arcsin\left[\frac{\gamma}{\sqrt{1+\gamma^2}}\, \sin\left(2 n \pi \, u\right)\right]
\label{psols-sech}
\ear
\begin{figure}[ht]
\centerline{\includegraphics[scale=0.9]{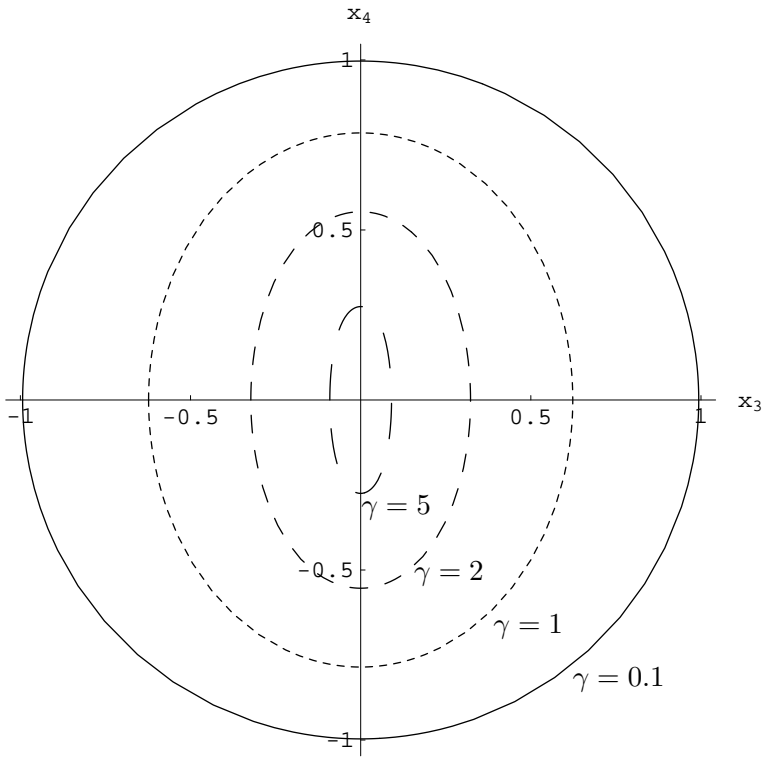}}
\begin{picture}(40,20)(60,-30)
\put(150,20){$\gamma=0.1$} \put(120,40){$\gamma=1$} \put(90,60){$\gamma=2$} \put(70,85){$\gamma=5$}
\end{picture}
\caption{Parametric plot of the stationary worldline instanton  paths (\protect{\ref{psols-sech}}) in the $(x_3, x_4)$ plane for the case of a time dependent electric field $E(t)=E\, {\rm sech}^2(\omega t)$. The paths are shown for various values of the adiabaticity parameter $\gamma=\frac{m\omega}{eE}$ defined in (\protect{\ref{ad}}), and $x_3$ and $x_4$ have been expressed in units of $\frac{m}{eE}$. Note that in the static limit,  $\gamma\to 0$, the instanton paths reduce to the circular ones of the constant field case shown in Figure \ref{fig1}.}
\label{fig2}
\end{figure}
These instanton paths are plotted in Figure 2 for various values of the adiabaticity parameter $\gamma$. In the static limit, when $\gamma\to 0$ with $\frac{\gamma}{\omega}\equiv \frac{m}{e E}$ fixed, we recover the circular stationary paths of the constant field case. In the short-pulse limit, $\gamma\to \infty$ with $\frac{m}{e E}$ fixed,  the paths become narrower  in the $x_3$ direction, and shrink in size.

To evaluate the stationary action $S_0$ we need $\dot{x}_4$:
\bear
\dot{x}_4(u)=a\, \frac{\cos (2\,n\,\pi \,u)}
  {
    {\sqrt{1 - \frac{\gamma ^2}
         {1 + \gamma ^2}\,{\sin^2 (2\,n\,\pi \,u)}}}}
\ear
Thus the stationary action $S_0$ is
\bear
S_0&=&m a  \int_0^1 du\, \frac{\cos^2(2 n \pi u)}{1-\frac{\gamma^2}{1+\gamma^2}\sin^2(2 n \pi u)}
\nn\\
&=&n\, \frac{m^2 \pi}{e E}\left(\frac{2}{1+\sqrt{1+\gamma^2}}\right)
\label{action-sech}\\\nn\\
&\sim&  \begin{cases}
{n\, \frac{m^2 \pi}{e E}\left(1-\frac{\gamma^2}{4}+\frac{\gamma^4}{8}+\dots\right)\quad , \quad \gamma\ll 1\cr
n\, \frac{2 m \pi}{\omega}\left(1-\frac{1}{\gamma}+\frac{1}{2\gamma^2}+\dots \right)\quad , \quad \gamma\gg 1}
\end{cases}
\label{action-sechapprox}
\ear
\begin{figure}[ht]
\centerline{\includegraphics[scale=0.9]{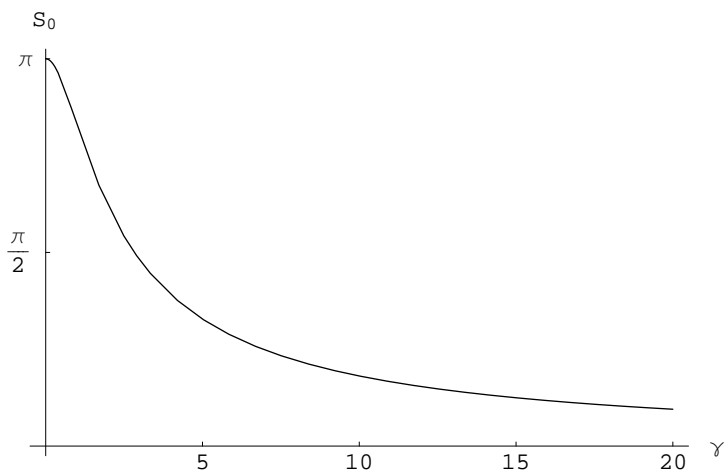}}
\begin{picture}(40,20)(60,-30)
\end{picture}
\caption{Plot of the instanton action $S_0$,  in units of $n\frac{m^2}{eE}$,  in (\protect{\ref{action-sech}}) for the time-dependent electric field $E(t)=E\, {\rm sech}^2(\omega t)$, plotted as a function of the adiabaticity parameter $\gamma$. Contrast this plot with the behavior in Figure \ref{fig7} for a spatial inhomogeneity of the same form.}
\label{fig3}
\end{figure}
This instanton action (\ref{action-sech}) is plotted in Figure \ref{fig3},  in units of $n\frac{m^2}{eE}$, as a function of the adiabaticity parameter $\gamma$.
Note that as $\gamma\to 0$, we recover the familiar instanton action of the constant field case. But as $\gamma$ increases, $S_0$ decreases, which means that the pair production rate is locally enhanced, relative to the locally constant field approximation with a field of the same peak magnitude. This is in full agreement with the WKB results \cite{popov}.

\subsection{Sinusoidal time dependent electric background : $E(t)=E\, \cos(\omega\,t)$.}
\label{tcos}

The Minkowski space time dependent electric field $E(t)=E\, \cos(\omega\,t)$ corresponds to the Euclidean space gauge field
\bear 
A_3(x_4)= - i\, \frac{E}{\omega}\,\sinh(\omega \, x_4) \quad .
\ear
Then the stationary $x_4(u)$ is determined by integrating (\ref{x4}), using the adiabaticity parameter $\gamma$ as defined in (\ref{ad}),
\bear
du=\frac{1}{a}\frac{d x_4}{\sqrt{1-\frac{1}{\gamma^2}\sinh^2(\omega \, x_4)}}
\ear
This has the solution
\bear
x_4=\frac{1}{\omega}\, {\rm arcsinh}\left[\frac{\gamma}{\sqrt{1+\gamma^2}}\, {\rm sd}\left(\frac{\sqrt{1+\gamma^2}}{\gamma}\, \omega \, a\, u\, {\Bigg |}  \frac{\gamma^2}{1+\gamma^2}\right)\right]
\ear
Here ${\rm sd}(x|\nu)$ is the Jacobi elliptic function \cite{abramowitz} with real elliptic parameter $0\leq \nu\leq 1$.
Then the stationary $x_3(u)$ is determined by integrating (\ref{x3}):
\bear
\frac{d x_3}{du}=-a\, \frac{1}{\sqrt{1+\gamma^2}}\, {\rm sd}\left(\frac{\sqrt{1+\gamma^2}}{\gamma}\, \omega \, a\, u\, {\Bigg |}  \frac{\gamma^2}{1+\gamma^2}\right)
\ear
This has the solution
\bear
x_3=\frac{1}{\omega}\, {\rm arcsin}\left[\frac{\gamma}{\sqrt{1+\gamma^2}}\, {\rm cd}\left(\frac{\sqrt{1+\gamma^2}}{\gamma}\, \omega \, a\, u\, {\Bigg |}  \frac{\gamma^2}{1+\gamma^2}\right)\right]
\ear
It is straightforward to verify that $\dot{x}_3^2+\dot{x}_4^2=a^2$ is satisfied.
Demanding that the solutions be periodic in $u$ with period 1 fixes the constant $a$ to be
\bear
a=\frac{\gamma}{\omega\, \sqrt{1+\gamma^2}}\,4\,{\bf K}\left(\frac{\gamma^2}{1+\gamma^2}\right)  n\quad , \quad n\in {\bf Z}^+
\label{acos}
\ear
where ${\bf K}(\nu)$ is the complete elliptic integral, which is the real quarter-period of the Jacobi elliptic functions \cite{abramowitz}. In this inhomogeneous case, the approximation condition (\ref{large}) becomes
\bear
\frac{m^2}{e E} \frac{4{\bf K}\left(\frac{\gamma^2}{1+\gamma^2}\right) n}{\sqrt{1+\gamma^2}} \gg 1
\eea
As in the constant field case (\ref{weakfield1}), this is a weak-field condition, although it also includes the adiabaticity parameter $\gamma$, which cannot be too large for a given peak field $E$.

\begin{figure}[h]
\centerline{\includegraphics[scale=0.9]{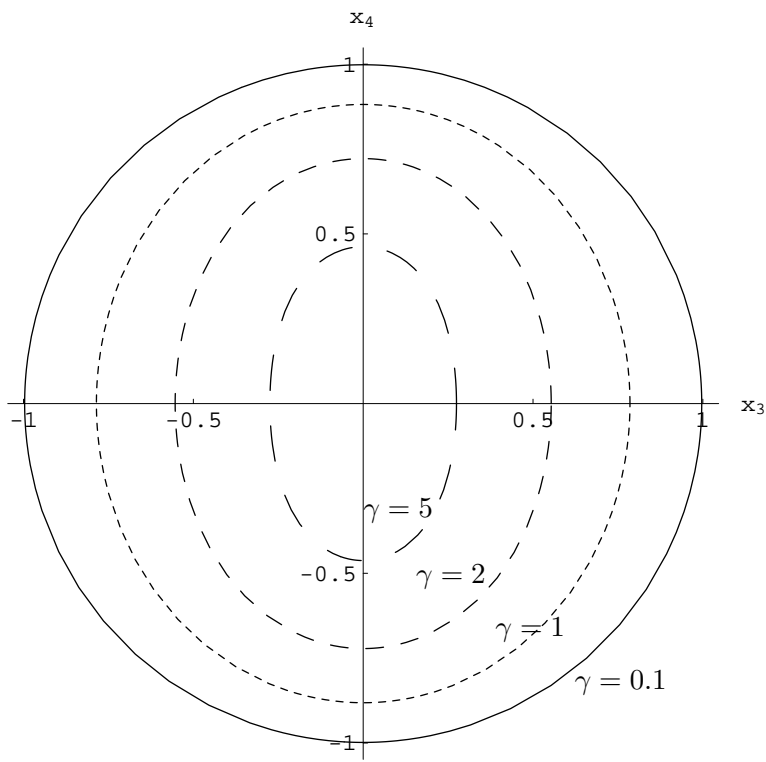}}
\begin{picture}(40,20)(60,-30)
\put(150,20){$\gamma=0.1$} \put(120,40){$\gamma=1$} \put(90,60){$\gamma=2$} \put(70,85){$\gamma=5$}
\end{picture}
\caption{Parametric plot of the stationary  worldline instanton paths (\protect{\ref{psols-cos}}) in the $(x_3, x_4)$ plane for the case of a time dependent electric field $E(t)=E\, \cos(\omega t)$. The paths are shown for various values of the adiabaticity parameter $\gamma=\frac{m\omega}{eE}$ defined in (\protect{\ref{ad}}), and $x_3$ and $x_4$ have been expressed in units of $\frac{m}{eE}$. Note that in the static limit,  $\gamma\to 0$, the instanton paths reduce to the circular ones of the constant field case shown in Figure \ref{fig1}. }
\label{fig4}
\end{figure}

The periodic stationary worldline instanton paths are:
\bear
x_3(u)&=&\frac{1}{\omega}\, {\rm arcsin}\left[\frac{\gamma}{\sqrt{1+\gamma^2}}\, {\rm cd}\left(4\,n\, {\bf K}\left(\frac{\gamma^2}{1+\gamma^2}\right) u\, {\Bigg |}  \frac{\gamma^2}{1+\gamma^2}\right)\right]
\\
x_4(u)&=&\frac{1}{\omega}\, {\rm arcsinh}\left[\frac{\gamma}{\sqrt{1+\gamma^2}}\, {\rm sd}\left(4\,n\, {\bf K}\left(\frac{\gamma^2}{1+\gamma^2}\right) u\, {\Bigg |}  \frac{\gamma^2}{1+\gamma^2}\right)\right]
\label{psols-cos}
\ear
These stationary instanton paths are plotted in Figure \ref{fig4} for various values of the adiabaticity parameter $\gamma$. In the static limit, when $\gamma\to 0$ with $\frac{\gamma}{\omega}\equiv \frac{m}{e E}$ fixed, we recover the circular stationary paths of the constant field case. In the high frequency limit, the instanton paths shrink in size and become narrower in the $x_3$ direction, a behavior that is qualitatively similar to the single-pulse case depicted in Figure \ref{fig2}.

\begin{figure}[h]
\centerline{\includegraphics[scale=0.9]{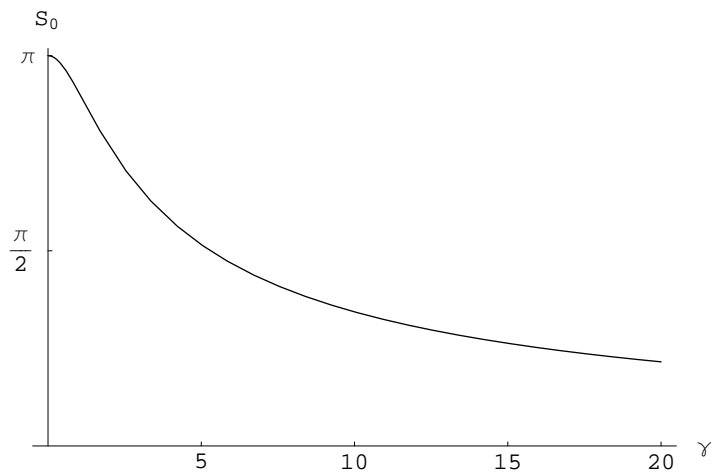}}
\begin{picture}(40,20)(60,-30)
\end{picture}
\caption{Plot of the instanton action $S_0$, in units of $n\frac{m^2}{eE}$, for the time-dependent electric field $E(t)=E\, \cos(\omega t)$, plotted as a function of the adiabaticity parameter $\gamma$. Contrast this plot with the behavior in Figure \ref{fig10} for a spatial inhomogeneity of the same form.}
\label{fig5}
\end{figure}

To evaluate the stationary action $S_0$ we need $\dot{x}_4$:
\bear
\dot{x}_4(u)=a\, {\rm cd}\left(4\,n\, {\bf K}\left(\frac{\gamma^2}{1+\gamma^2}\right) u\, {\Bigg |}  \frac{\gamma^2}{1+\gamma^2}\right)
\ear
Thus the stationary action $S_0$ is
\bear
S_0&=&m\, a\,  \int_0^1 du\, {\rm cd}^2\left(4\,n\, {\bf K}\left(\frac{\gamma^2}{1+\gamma^2}\right) u\, {\Bigg |}  \frac{\gamma^2}{1+\gamma^2}\right)
\nn\\ \nn\\
&=&n\, \frac{m^2 }{e E}\, \frac{4\sqrt{1+\gamma^2}}{\gamma^2}\left[{\bf K}\left(\frac{\gamma^2}{1+\gamma^2}\right)-{\bf E}\left(\frac{\gamma^2}{1+\gamma^2}\right)\right]
\label{action-cos}\\ \nn\\
&\sim& \begin{cases}
{n\, \frac{m^2 \pi}{e E}\left(1-\frac{\gamma^2}{8}+\frac{3\gamma^4}{64}+\dots\right)\quad , \quad \gamma\ll 1\cr
n\, \frac{4 m}{\omega}\left(\ln(4 \gamma)-1+\frac{\ln(4 \gamma)}{4\gamma^2}+\dots \right)\quad , \quad \gamma\gg 1}
\end{cases}
\label{log}
\ear
This instanton action is plotted in Figure \ref{fig5},  in units of $n\frac{m^2}{eE}$, as a function of the adiabaticity parameter $\gamma$.
In the static limit,  $\gamma\to 0$ with $\frac{\gamma}{\omega}\equiv \frac{m}{e E}$ fixed, we recover the familiar instanton action of the constant field case.
In the high frequency limit, $S_0$ decreases monotonically, which means that the pair production rate is locally enhanced. This behavior is qualitatively similar to the single-pulse case considered in the previous subsection, and plotted in Figure \ref{fig4}. It is also in full agreement with the WKB results \cite{brezin,popov}.

Notice that in the large $\gamma$ limit of the sinusoidal case, the logarithmic behavior shown in (\ref{log}) means that the leading exponential part of the pair production rate is
\bear
\Gamma\sim e^{-S_0}
\sim \left(\frac{e E}{m\omega}\right)^{\frac{4 n m}{\omega}}
\label{pert}
\ear
This has a perturbative multi-photon form, being $\left(\frac{e A}{m}\right)^2$ raised to the number of photons of frequency $\omega$ needed to make up the pair production energy $2m$. Thus, remarkably, the instanton action (\ref{action-cos}) interpolates between the nonperturbative constant-field form (\ref{action-constant}) and the pertubative multi-photon form (\ref{pert}) as the adiabaticity parameter $\gamma$ goes from $0$ to $\infty$. This was noted in WKB analyses \cite{popov,brezin}, and generalizes to a relativistic context behavior first found by Keldysh \cite{keldysh} in WKB studies of ionization processes.
The large $\gamma$ limit of the $E(t)=E\, {\rm sech}^2(\omega t)$ case, shown in (\ref{action-sechapprox}), is also perturbative, as it has a simple expansion in $E$. However, it is not as obviously of a multi-photon form such as (\ref{pert}), since the background is not monochromatic. 

\section{Spatially Inhomogeneous Electric Fields}
\label{space}

An appealing feature of the worldline instanton approach is that it may be applied also to spatially inhomogeneous electric fields. The instanton condition is still given by (\ref{statcond}). Consider a spatially inhomogeneous electric field directed along the (negative) $x_3$ direction in Minkowski space. In Euclidean space, we choose a gauge in which the only nonzero component is $A_4$, which is a function of $x_3$ only:
\bear
A_4=A_4(x_3) \quad ; \quad A_\mu =0\quad {\rm for}\quad \mu\neq 4
\label{spatiala}
\ear
Then the instanton equations (\ref{coupled1}), (\ref{coupled2}) reduce to
\bear
\dot{x}_3&=&a\sqrt{1+\left(\frac{e A_4(x_3)}{m}\right)^2}
\\
\dot{x}_4&=&-\frac{i e a}{m}\, A_4(x_3)
\label{spatialinstanton}
\ear
which are very similar in form to the worldline instanton equations (\ref{x3}) and (\ref{x4}) for a time dependent electric field. Similarly, when evaluated on a worldline instanton path, the stationary action is
\bear
S_0&=&m\sqrt{ \int_0^1du\, \dot{x}^2}+i e\int_0^1 du\,\frac{dx_4}{du}\, A_4(x_3)\non\\
&=& \frac{m}{a}\int_0^1du\, (\dot{x}_3)^2
\label{s0x}
\ear

\subsection{Constant electric field}
\label{xconstant}

For a constant electric field, we could have chosen the static gauge, $A_4=-i E x_3$, in which case we find instanton paths
\bear
x_3(u)&=&\frac{m}{e E} \sin(2 \pi n u)\nn\\
x_4(u)&=&\frac{m}{e E} \cos(2 \pi n u)
\ear
which are once again circular paths of radius $\frac{m}{e E}$. And, as before, the stationary action is $S_0=n\frac{m^2 \pi}{e E}$, for instanton number $n$.

\subsection{Single-bump electric field : $E(x_3)=E\, {\rm sech}^2(k x_3)$}
\label{xsech}

\begin{figure}[ht]
\centerline{\includegraphics[scale=0.9]{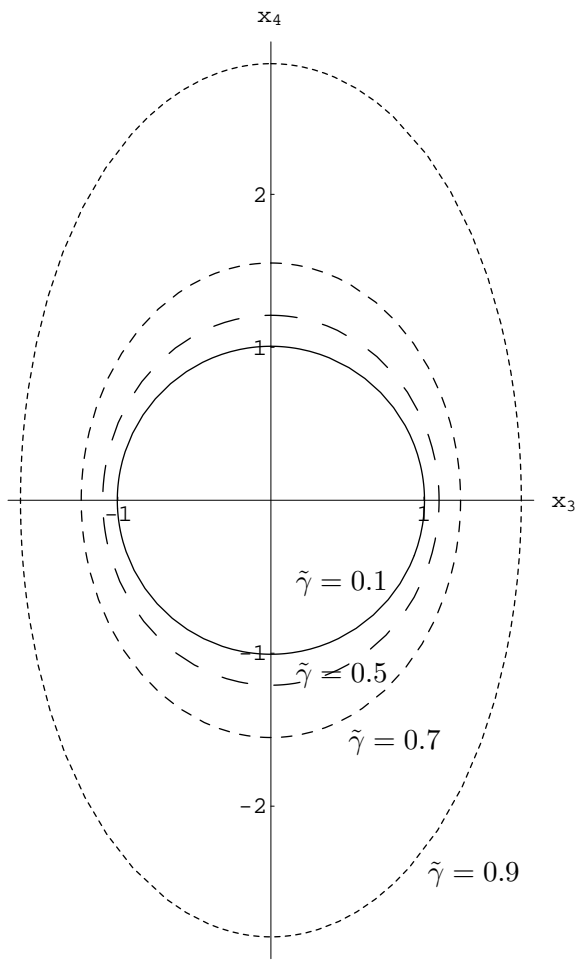}}
\begin{picture}(40,20)(60,-30)
\put(80,140){$\tg=0.1$} \put(80,105){$\tg=0.5$} \put(100,80){$\tg=0.7$} \put(130,30){$\tg=0.9$}
\end{picture}
\caption{Instanton paths for the spatially inhomogeneous electric field $E(x)=E\, {\rm sech}^2(k x)$ for various values of the inhomogeneity parameter $\tg$ defined in (\protect{\ref{tgamma}}). As $\tg\to 0$ we recover the circular paths of the constant field case, but as $\tg\to 1$ the loops become infinitely large.}
\label{fig6}
\end{figure}

Now consider a time-independent but spatially inhomogeneous electric field with a single-bump profile of the form $E(x)=E\, {\rm sech}^2(k x)$. This case is also soluble in the sense that the pair production rate can be expressed as an integral over momenta of a known function \cite{nikishov2}. It has also recently been analyzed numerically in the numerical worldline approach \cite{giesklingmuller}. This case can be realized with a gauge field 
\bear
A_4=-i \frac{E}{k} {\rm tanh}(k x_3)
\ear
where $1/k$ characterizes the length scale of the spatial inhomogeneity. The instanton equations are
\bear
\dot{x}_3 &=& a\sqrt{1-\frac{1}{\tilde{\gamma}^2}\, {\rm tanh}^2(k x_3)}\nn\\
\dot{x}_4&=& -\frac{a}{\tilde{\gamma}}\,  {\rm tanh}(k x_3)
\label{sechpath}
\ear
where we have defined an inhomogeneity parameter
\bear
\tilde{\gamma}=\frac{m k}{e E}\quad ,
\label{tgamma}
\ear 
in analogy with Keldysh's adiabaticity parameter (\ref{ad}). It is easy to verify that the periodic worldline instanton paths are:
\bear
x_3(u)&=& \frac{m}{e E} \frac{1}{\tilde{\gamma}} \, {\rm arcsinh}\left(\frac{\tg}{\sqrt{1-\tg^2}}\, \sin (2 \pi n u)\right)\nn\\
x_4(u)&=&  \frac{m}{e E} \frac{1}{\tilde{\gamma}\sqrt{1-\tg^2}} \, {\rm arcsin}\left(\tg\, \cos (2 \pi n u)\right)
\ear
These instanton loops are plotted in Figure \ref{fig6} for various values of the inhomogeneity parameter $\tg$. Note that they reduce once again to the circular constant field loops in the constant limit $\tg\to 0$, but as $\tg$ approaches 1 the loops grow in size, becoming very elongated in the $x_4$ direction. Recall \cite{nikishov2,giesklingmuller} that there is a cutoff value of $k$ beyond which there is no pair creation, and this cutoff value corresponds to $\tg=1$. Physically, this is where the spatial width of the electric field is smaller than the electron Compton wavelength, at which point a virtual pair is not able to extract enough energy from the electric field to become real asymptotic particles. The worldline instanton approach naturally incorporates this cutoff as the instanton paths are real only for $\tg<1$. 

\begin{figure}[ht]
\centerline{\includegraphics[scale=0.9]{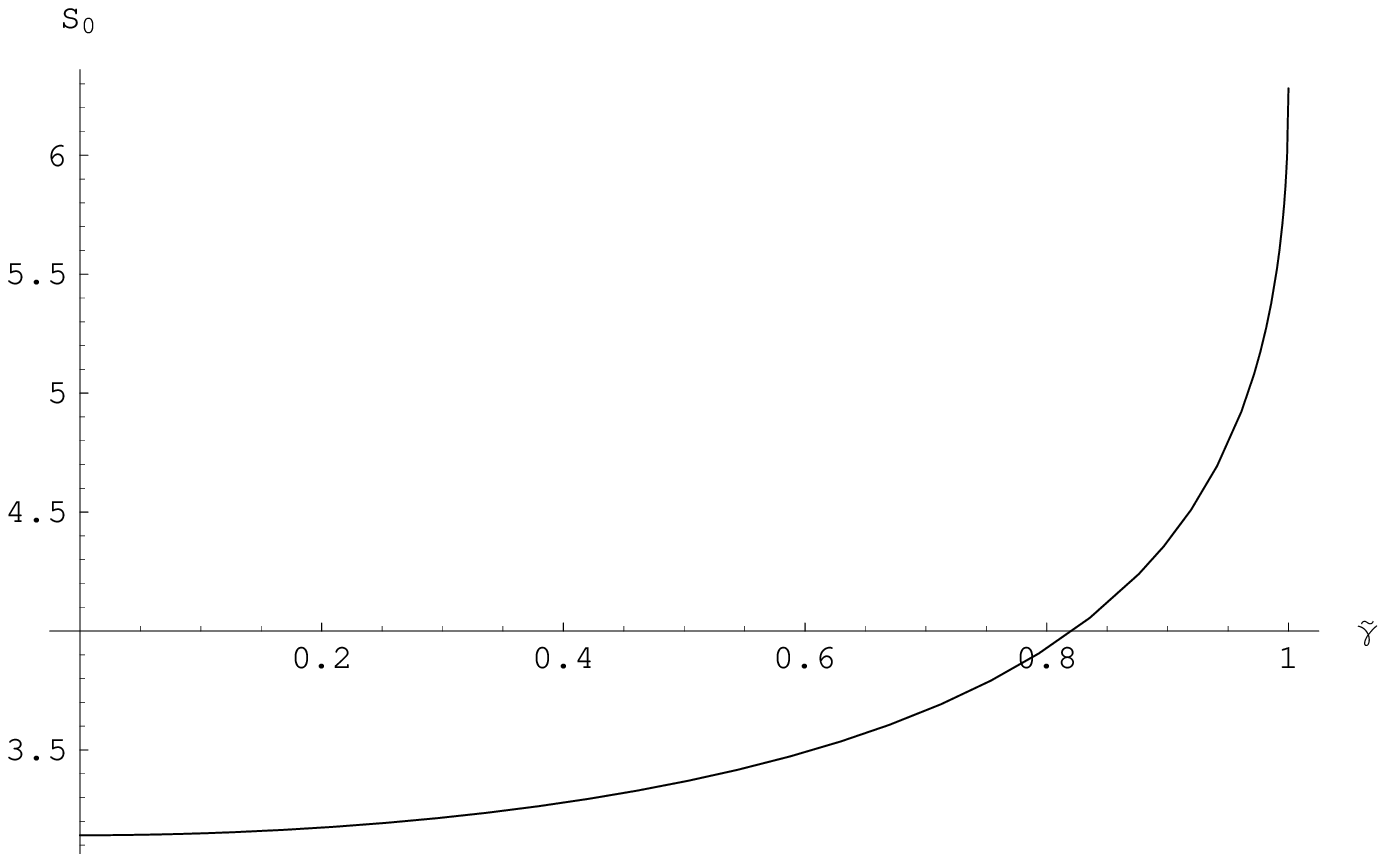}}
\caption{The stationary action $S_0$, in units of $n\frac{m^2}{e E}$, plotted as a function of the inhomogeneity parameter $\tg$ defined in (\protect{\ref{tgamma}}). Note that $S_0$ increases with $\tg$ and diverges as $\tg\to 1$. Contrast this plot with the behavior in Figure \ref{fig3} for a temporal inhomogeneity of the same form.}
\label{fig7}
\end{figure}

Given the instanton path (\ref{sechpath}), the stationary action is
\bear
S_0 &=&n\, \frac{m^2 \pi}{e E}\left(\frac{2}{1+\sqrt{1-\tg^2}}\right)
\label{action-sechx}
\ear
This stationary action is plotted in Figure \ref{fig7}, in units of $n \frac{m^2 \pi}{e E}$, as a function of the inhomogeneity parameter $\tg$. 

Observe that this spatially inhomogeneous case is closely related to the temporally inhomogeneous case considered in section \ref{tsech}. Indeed, they are related by the analytic continuation $\gamma\to i\tg$, as can be seen in the expressions for the instanton paths and in the expressions for the stationary actions. This analytic continuation is similar to one connecting time dependent electric fields with spatially inhomogeneous magnetic backgrounds \cite{dhelectric,fry}. But we note here that as $\tg$ increases from 0, the stationary action $S_0$ grows, and becomes infinite as $\tg\to 1$. This is precisely the opposite of the behavior in the temporally inhomogeneous case. So, it appears that while a temporal inhomogeneity locally enhances the pair production, a spatial inhomogeneity locally decreases the pair production rate. Correspondingly, the behavior is also opposite  for the instanton paths: as $\tg$ grows from 0 the paths grow from the circular constant field instanton paths (see Figure \ref{fig7}), while in the temporally inhomogeneous case as $\gamma$ grows from 0 the instantons loops shrink from the circular constant field instanton paths (see Figure \ref{fig3}). 

\begin{figure}[ht]
\centerline{\includegraphics[scale=0.9]{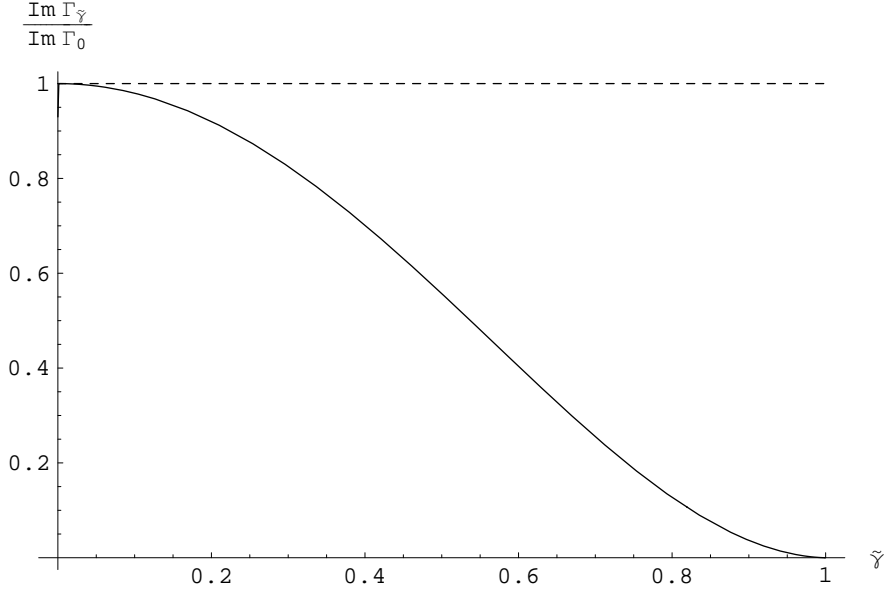}}
\begin{picture}(40,20)(60,-30)
\end{picture}
\caption{The ratio (\protect{\ref{ratio}}) of ${\rm Im} \Gamma$ for the spatially inhomogeneous electric field $E(x)=E\, {\rm sech}^2(k x)$ to the locally constant field result, in the single instanton loop approximation, plotted as a function of the inhomogeneity parameter $\tg$ defined in (\protect{\ref{tgamma}}).}
\label{fig8}
\end{figure}

Finally, including the WKB prefactor, the ratio of the pair production rate in the inhomogeneous field relative to the locally constant field value (obtained by substituting the inhomogeneous field in the constant field answer and integrating) is given by the ratio of the imaginary parts of the effective action in each case: 
\bear
\frac{{\rm Im} \Gamma_{\tg} }{{\rm Im} \Gamma_0} = \left(1-\tg^2 \right)^{5/4} \, \exp \left[- \frac{m^2 \pi}{e E}\left(\frac{2}{1+\sqrt{1-\tg^2}} -1\right)\right]
\label{ratio}
\ear 
This is plotted in Figure \ref{fig8}, and it compares very accurately with the numerically integrated exact result of Nikishov \cite{nikishov2} and with the recent numerical results of Gies et al (see Figure 3 in \cite{giesklingmuller}). 

\subsection{Spatially sinusoidal electric field : $E(x)=E\, \cos(k x)$}
\label{xcos}

A similar analysis can be done for a spatially inhomogeneous electric field of sinusoidal form. This case can be realized with a gauge field 
\bear
A_4=-i \frac{E}{k} {\rm sin}(k x_3)
\ear
where $1/k$ characterizes the length scale of the spatial inhomogeneity. The instanton  paths are:
\bear
x_3(u)&=& \frac{m}{e E} \frac{1}{\tilde{\gamma}} \, {\rm arcsinh}\left[\frac{\tg}{\sqrt{1-\tg^2}}\, {\rm cd}\left(4\,n\, {\bf K}\left(-\frac{\tg^2}{1-\tg^2}\right) u\, {\Bigg |}  -\frac{\tg^2}{1-\tg^2}\right)\right]\nn\\
x_4(u)&=& \frac{m}{e E} \frac{1}{\tilde{\gamma}} \, {\rm arcsin}\left[\frac{\tg}{\sqrt{1-\tg^2}}\, {\rm sd}\left(4\,n\, {\bf K}\left(-\frac{\tg^2}{1-\tg^2}\right) u\, {\Bigg |}  -\frac{\tg^2}{1-\tg^2}\right)\right]
\label{cospath}
\ear

\begin{figure}[ht]
\centerline{\includegraphics[scale=0.9]{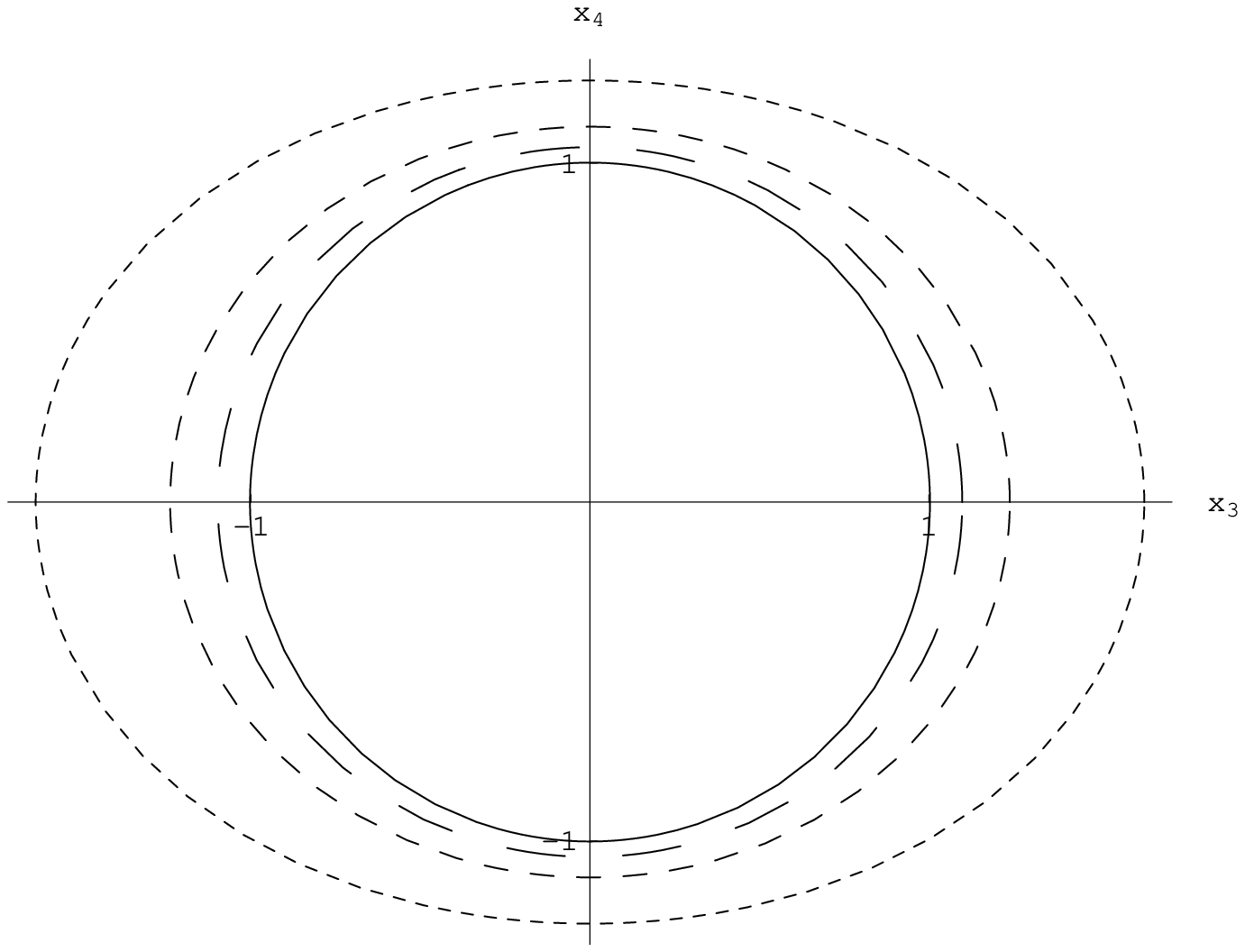}}
\begin{picture}(40,20)(60,-30)
\put(100,55){$\tg=0.1$} \put(160,85){$\tg=0.5$} \put(170,50){$\tg=0.7$} \put(180,20){$\tg=0.9$}
\end{picture}
\caption{Instanton paths for the spatially inhomogeneous electric field $E(x)=E\, \cos(k x)$ for various values of the inhomogeneity parameter $\tg$ defined in (\protect{\ref{tgamma}}). As $\tg\to 0$ we recover the circular paths of the constant field case, but as $\tg\to 1$ the loops become elongated in the $x_3$ direction.}
\label{fig9}
\end{figure}

These instanton loops are plotted in Figure \ref{fig9} for various values of the inhomogeneity parameter $\tg$. Note that they reduce once again to the circular constant field loops in the constant limit $\tg\to 0$, but as $\tg$ approaches 1 the loops grow in size, becoming very elongated in the $x_3$ direction, and flattening off to the constant values $\pm \frac{\pi}{2}$ in the $x_4$ direction. Once again, there is a cutoff value of $k$ beyond which there is no pair creation, and this cutoff value corresponds to $\tg=1$.  

\begin{figure}[ht]
\centerline{\includegraphics[scale=0.9]{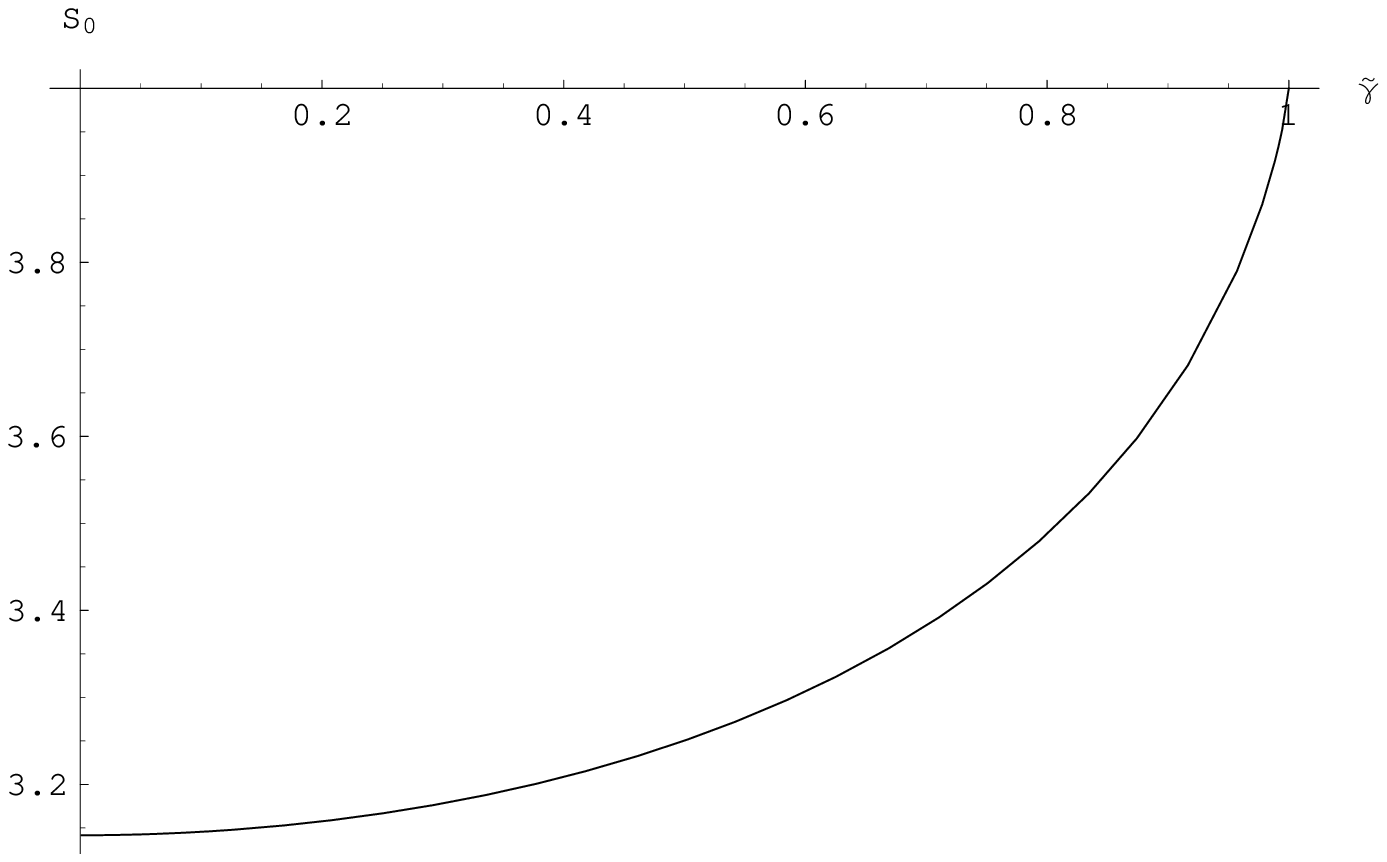}}
\caption{The stationary action $S_0$, in units of $n \frac{m^2}{e E}$, plotted as a function of the inhomogeneity parameter $\tg$ defined in (\protect{\ref{tgamma}}). Note that $S_0$ increases with $\tg$ and approaches $4$ as $\tg\to 1$. Contrast this plot with the behavior in Figure \ref{fig5} for a temporal inhomogeneity of the same form.}
\label{fig10}
\end{figure}

\begin{figure}[ht]
\centerline{\includegraphics[scale=0.9]{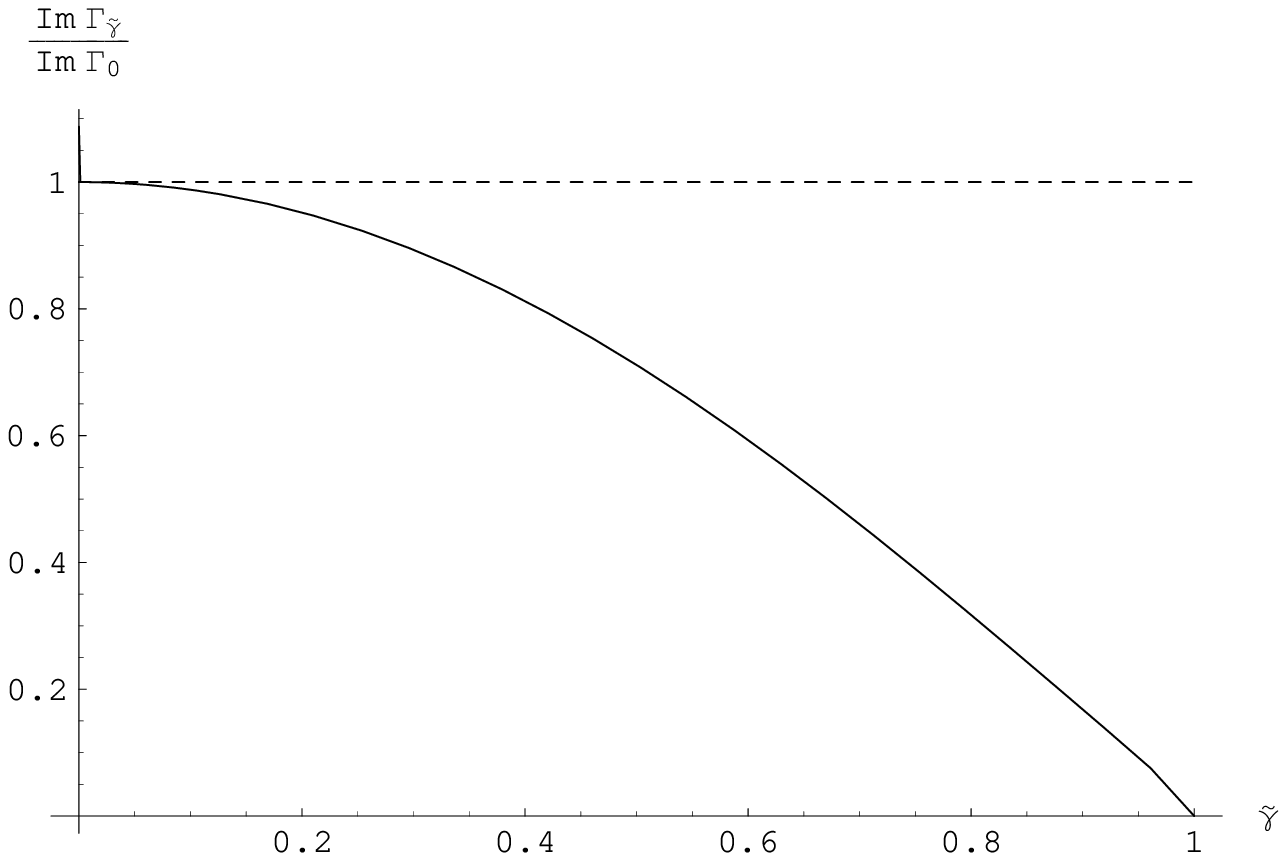}}
\caption{The ratio (\protect{\ref{ratio}}) of $Im \Gamma$ for the spatially inhomogeneous electric field $E(x)=E\, \cos(k x)$ to the constant field result, in the single instanton loop approximation, plotted as a function of the inhomogeneity parameter $\tg$ defined in (\protect{\ref{tgamma}}).}
\label{fig11}
\end{figure}

Given the instanton path (\ref{cospath}), the stationary action is
\bear
S_0&=&n\, \frac{m^2 }{e E}\, \frac{4\sqrt{1-\tg^2}}{\tg^2}\left[{\bf E}\left(-\frac{\tg^2}{1-\tg^2}\right)-{\bf K}\left(-\frac{\tg^2}{1-\tg^2}\right)\right]
\label{action-cosx}\\ \nn\\
&\sim& \begin{cases}
{n\, \frac{m^2 \pi}{e E}\left(1+\frac{\tg^2}{8}+\frac{3\tg^4}{64}+\dots\right)\quad , \quad \tg\ll 1\cr
n\,  \frac{m^2 }{e E}\left(4+2 \left(1-\tg\right)\left[3+\log\left(\frac{1-\tg}{8}\right)\right]+\dots \right)\quad , \quad \tg\to 1}
\end{cases}
\label{xcosaction}
\ear
This stationary action is plotted in Figure \ref{fig10}, in units of $n \frac{m^2 \pi}{e E}$, as a function of the inhomogeneity parameter $\tg$. 
As in the spatially inhomogeneous case considered in Section \ref{xsech}, the instanton paths grow with the inhomogeneity parameter $\tg$, and the instanton action also grows, which means that the pair production is supressed by the increasing inhomogeneity. Finally, including the WKB prefactor, our worldline instanton result for the ratio of the pair production rate relative to the locally constant field value is
\bear
\frac{{\rm Im} \Gamma_{\tg} }{{\rm Im} \Gamma_0} = \frac{\pi^{3/2} \tg  (1-\tg^2)^{3/4}}{4 {\bf K}\sqrt{{\bf E}-{\bf K}}} \, \exp \left[-\frac{m^2\pi}{e E} \left(\frac{4\sqrt{1-\tg^2}}{\tg^2}\left[{\bf E}-{\bf K}\right]
-1\right) \right]
\label{cosratio}
\ear
where ${\bf E}\equiv{\bf E}\left(\frac{-\tg^2}{1-\tg^2}\right)$, and ${\bf K}\equiv{\bf K}\left(\frac{-\tg^2}{1-\tg^2}\right)$. 
This ratio is plotted in Figure \ref{fig11}. 

\section{Worldline Instantons for Spinor QED}
\label{spinor}

In this Section we show that all our considerations can be easily generalized from the scalar to the spinor loop case. In fact, we will show that the worldline instanton for the spinor case is identical to the worldline instanton for the scalar case.

Recall that the path integral representation of the one-loop effective action due to a spin half particle in the loop differs from (\ref{PI}) only by a global factor of $-\half$ and the insertion of the following `spin factor' $S[x,A]$ under the path integral \cite{feynman},
\bear
S[x,A] = \tr_{\Gamma}{\cal P}\,\,\e^{{i\over 2}e\sigma^{\mn}\int_0^Td\tau F_{\mn}(x(\tau))}
\label{defspinfactor}
\ear   
Here $\tr_{\Gamma}$ denotes the Dirac trace and ${\cal P}$ the path ordering operator. 
For the special cases considered in the present paper the path ordering has no effect,
since the $F_{\mn}(x(\tau))$ at different proper-times commute. The spin factor
then immediately reduces to
\bear
S[x,A] = 4\,\cos\Bigl[e\int_0^T d\tau\, E(x(\tau))\Bigr]
= 4\,\cos\Bigl[eT\int_0^1 du\, E(x(u))\Bigr]
\label{simpspinfactor}
\ear
where $E(x(u))$ is the electric field evaluated on the path $x(u)$.

Our analysis in this paper has involved doing the $T$ integral in (\ref{PI}) first, and then making a semiclassical instanton approximation for the remaining path integral in (\ref{bessel}). Recall that (\ref{bessel}) is obtained by evaluating the $T$ integral at the critical point (\ref{critical}). The key observation is that in Euclidean space the exponent in the spin factor (\ref{simpspinfactor}) is imaginary. Therefore the spin factor does not affect the determination of the stationary point $T_0$, nor does it modify the stationary (worldline instanton) equations (\ref{statcond}). The only effect of the spin factor is that we need to include the factor (\ref{simpspinfactor}) evaluated on the worldline instanton path. 

For example, in the constant field case, the spin factor is trivially
\bear
4\cos\Bigl[ e T_0 E\Bigr] =4\cos(\pi n)=4(-1)^n
\label{constantspin}
\ear
where we have used the expression (\ref{aperiod}) for the parameter $a$ in the constant field case. Including this additional alternating sign factor, along with the global factor of $-\frac{1}{2}$, we see that the imaginary part of the effective Lagrangian is
modified from the scalar QED form in (\ref{L1scalim}) to the well-known spinor QED form \cite{schwinger}:
\bear
{\rm Im}\,{\cal L}_{\rm spin}[E] =
\frac{e^2 E^2}{8\pi^3}\, \sum_{n=1}^\infty \frac{1}{n^2}
\,\exp\left[-\frac{m^2 \pi n}{e E}\right]
\label{L1spinim}
\end{eqnarray}

%
%
%

Remarkably, this simple alternating sign expression (\ref{constantspin}) for the spin factor applies not just for the constant field case, but actually extends to our more general class of inhomogeneous background electric fields where the single nonzero component of the gauge field is either $A_3(x_4)$ or $A_4(x_3)$. Consider, for example, the time-dependent electric field considered in Section \ref{tsech}, which has Minkowski electric field $E(t)=E{\rm sech}^2(\omega t)$. This corresponds to a Euclidean electric field $E(x_4)=E\sec^2(\omega x_4)$. Evaluating $E(x_4(u))$ on the worldline instanton solution (\ref{psols-sech}), and recalling the expression (\ref{asech}) for the parameter $a$, we find that the argument of the spin factor is
\bear
\frac{e  a}{2 m} \int_0^1 du\, E(x_4(u))&=&\pi n \sqrt{1+\gamma^2}\int_0^1 \frac{du}{1+\gamma^2 \cos^2(2\pi n u)}\nonumber \\
&=&\pi n
\ear
Similarly, consider the time-dependent electric field considered in Section \ref{tcos}, which has Minkowski electric field $E(t)=E\,\cos(\omega t)$. This corresponds to a Euclidean electric field $E(x_4)=E \cosh(\omega x_4)$. Evaluating $E(x_4(u))$ on the worldline instanton solution (\ref{psols-cos}), and recalling the expression (\ref{acos}) for the parameter $a$, we find that the argument of the spin factor is
\bear
\frac{e  a}{2 m} \int_0^1 du\, E(x_4(u))&=&2 n \frac{{\bf K}\left(\frac{\gamma^2}{1+\gamma^2}\right)}{ \sqrt{1+\gamma^2}} \int_0^1 du \sqrt{1+\frac{\gamma^2}{1+\gamma^2} \,{\rm sd}^2\left( 4 n {\bf K}\left(\frac{\gamma^2}{1+\gamma^2}\right) u\, \Bigg | \frac{\gamma^2}{1+\gamma^2}\right)}
\nonumber \\
&=&\pi n
\ear
To see that this holds in general, we use the periodicity of the worldline instanton solution to restrict the interval of the $u$ integral to a single interval on which $\dot{x}_4$ has the same sign. For a worldline instanton path of winding number $n$, there are $2n$ such intervals, and each contributes the same amount to $ \int_0^1 du\, E(x_4(u))$. Without loss of generality we take the sign of $\dot{x}_4$ to be positive on this interval. On this interval we use (\ref{x4}) and change integration variable from $u$ to $y=i e A_3/m$, noting from (\ref{x4}) that $\dot{x}_4$ vanishes at $y=\pm 1$ for the instanton path. Thus
\bear
\frac{e a}{2 m}\int_0^1du E(x_4(u))&=& {iea\over 2m} \int_0^1 du 
\frac{d A_3}{dx_4}\Bigg |_{x_4(u)} \non\\
&=& 2n\left( \half\int_{-1}^1 \frac{dy}{\sqrt{1-y^2}} \right)\non\\ 
&=& n\pi
\label{calcspinfactor}
\ear

A similar analysis applies for the spatially inhomogeneous electric fields with $A_4=A_4(x_3)$, which were discussed in Section \ref{space} for scalar QED. The implication is that for spinor problems we find the worldline instanton just as in the scalar case, and simply include a factor of $-2(-1)^n$ inside the sum over winding numbers of the instanton paths. We note that this result is consistent with Nikishov's virial representation \cite{lebedev} of the imaginary part of the effective action,
\bear
{\rm Im}\, \Gamma=\mp \frac{(2s+1)}{2}{\rm Tr}_p \ln \left(1\mp \bar{n}_p\right)
\label{virial}
\ear
where the upper/lower signs refer to the spinor/scalar case, $s$ is the spin, and $\bar{n}_p$ is the mean number of pairs produced in a state with momentum $p$.

\section{Conclusion}
\label{conclusions}

To conclude, we note that the worldline instanton technique gives the leading instanton action in agreement with the standard WKB approach \cite{brezin,popov}. 
For the classes of inhomogeneous background fields considered here, the instanton action can be expressed as
\bear
S_0=\cases{m\, a\, \int_0^1 du\, \left(1+\left(\frac{e\, A_3(x_4)}{m}\right)^2\right)\quad , \quad {\rm time\,\, dependent\,\, field}\cr 
m\, a\, \int_0^1 du\, \left(1+\left(\frac{e\, A_4(x_3)}{m}\right)^2\right)\quad , \quad {\rm space\,\, dependent\,\, field}}
\label{s0wkb}
\ear
Express the gauge field in the temporally or spatially dependent case as
\bear
A_3(x_4)=-i\, \frac{E}{\omega}\, f(\omega\, x_4)\quad , \quad A_4(x_3)=-i\, \frac{E}{k}\, f(k\, x_3)
\label{a3f}
\ear
for some function $f$. Then
\bear
S_0=\frac{4 m^2 n}{eE} \int_{0}^1 dy\, \frac{\sqrt{1-y^2}}{|f^\prime|}
\label{wkb}
\ear
where $y=\frac{1}{\gamma} f(v)$ or $y=\frac{1}{\tg} f(v)$, respectively, and where $f'(v)$ is to be re-expressed as a function of $y$. This expression (\ref{wkb}) agrees with the WKB expression \cite{popov,brezin} for the exponential part of the pair production rate in an electric field of the form (\ref{gauge}) or (\ref{spatiala}), respectively. The unified worldline instanton approach
gives an interesting spacetime picture of the dependence of the pair production on the form of the inhomogeneity. The worldline instanton loops shrink with increasing temporal inhomogeneity of the background electric field, and expand with increasing spatial inhomogeneity of the background electric field. Correspondingly, the local pair production, compared to a locally constant field of the same peak magnitude, is increased or suppressed, respectively. 

Several important issues remain and will be addressed in future work. First, the full pair production rate involves also a determinant prefactor corresponding to the contribution of fluctuations about the stationary worldline instanton path(s). This prefactor can be deduced from the WKB approach \cite{popov,brezin}, as was done for the plots in Figures \ref{fig8} and \ref{fig11}, but it can also be computed directly in the worldline approach \cite{ds-flucs}. In the constant $E$ field case the fluctuation operator has simple eigenvalues and the determinant prefactor can be computed straightforwardly \cite{affleck}. 

Another important question is what can be said about higher loop effects. It is particularly interesting that the worldline instanton approach has the potential to address higher loops, while it is not at all clear how to address higher loops in the WKB language. The main result of Affleck et al's work \cite{affleck} is that for a constant $E$ field, the instanton approach provides a way to resum the leading effect of all higher loops in the situation where the constant field $E$ is weak, but the coupling $e$ is arbitrary. This is because the instanton solution remains a stationary point even after taking the additional interaction term into account which in the worldline formalism represents 
virtual photon exchanges in the loop. Although this property is presumably specific to the constant field
case, it would be very interesting if this type of analysis could be extended to the general worldline instanton loops for  inhomogeneous background fields. 
Some steps in this direction in formulating such a strong-coupling expansion were taken already in the the second paper of \cite{halpern}. Finally, Gies et al have recently shown \cite{giesklingmuller} how one can study the imaginary part of the worldline effective action using a Monte Carlo numerical method, summing over an ensemble of closed spacetime loops. The precise role of the worldline instanton loops in this Monte Carlo approach remains to be clarified, but the close agreement of Figure \ref{fig8} with the results in \cite{giesklingmuller} is suggestive.

\vskip 1cm

{\bf Acknowledgements:} We are grateful to Holger Gies for helpful comments and correspondence. GD also thanks the US DOE for support through the grant DE-FG02-92ER40716, and thanks Choonkyu Lee for discussions and comments.

\end{document}